\documentclass[final,3p,times]{elsarticle}
\usepackage{amsfonts}
\usepackage{amsmath}
\usepackage{amssymb}
\usepackage{graphicx}
\usepackage{color}
\usepackage{physics}
\usepackage{float}
\usepackage{mathtools}
\usepackage{algorithm}
\usepackage{algpseudocode}
\usepackage{wrapfig}
\usepackage{mathbbol}
\usepackage{upgreek}
\usepackage[normalem]{ulem}
\biboptions{sort&compress}

\usepackage[colorinlistoftodos]{todonotes}
\usepackage[colorlinks=true, allcolors=blue]{hyperref}
\usepackage{mwe}
\usepackage{epstopdf}
\usepackage[doipre={DOI:~}]{uri}

\renewcommand{\cite}{\citep}
\newcommand{\ii}{\textrm{i}}

\newcommand \kv {\mathbf{G}}

\newcommand{\Gpar}{G^\parallel} 
\newcommand{\Gperp}{G^\perp}

\newcommand{\epar}{\bar{\varepsilon}^\parallel} 
\newcommand{\eperp}{\bar{\varepsilon}^\perp}

\newcommand{\parall}{{||}}
\newcommand{\Gvn}{\mathbf{G}_j}
\newcommand{\Gv}{\mathbf{G}}

\newcommand{\rv}{\mathbf{r}}

\newcommand{\bv}{\mathbf{b}}

\newcommand{\I}{\mathbb{i}}

\begin{document}
\title{Modeling dislocations in quasicrystals through amplitude equations}

\author[TUD]{Marcello De Donno}
\author[OSLO]{Luiza Angheluta}
\author[TUD,DCMS]{Marco Salvalaglio\corref{cor}}
\ead{marco.salvalaglio@tu-dresden.de}
\cortext[cor]{Corresponding author}

\address[TUD]{Institute  of Scientific Computing,  TU  Dresden,  01062  Dresden,  Germany}
\address[OSLO]{Njord Centre, Department of Physics, University of Oslo, 0371 Oslo, Norway}
\address[DCMS]{Dresden Center for Computational Materials Science (DCMS), TU Dresden, 01062 Dresden, Germany}

\begin{abstract}
Quasicrystals (QCs) are a class of aperiodic ordered structures that emerge in various systems, from metallic alloys to soft matter and driven non-equilibrium systems. Within a mesoscale theory based on slowly-varying complex amplitudes for QCs, we track dislocations as topological defects harbored by the amplitudes and characterize their Burgers vectors and induced deformations. We study the formation of dislocations at semicoherent interfaces, particularly those emerging from rotated inclusions, and find a hierarchy of dislocations forming at such interfaces. We further analyze interfaces in strained systems, revealing conditions for the emergence of periodic dislocation arrays and discussing the energetics of dislocations associated with different phonon and phason deformations. The stability, interaction, and motion of dislocation dipoles and quadrupoles are also discussed. These findings provide new insights into the mesoscale modeling of dislocations in QCs and their distinct behavior compared to conventional crystals, while demonstrating a versatile framework for studying dislocations in systems exhibiting quasicrystalline order.
\end{abstract}


\maketitle

\section{Introduction}

Quasicrystals (QCs) are aperiodic structures characterized by discrete diffraction patterns. Initially explored as a mathematical concept \cite{penrose1974role,MACKAY1982609}, they were eventually discovered in metallic alloys in the 1980s \cite{Shechtman1984metallic,Levine1984quasicrystals}, prompting a redefinition of the very concept of crystal \cite{:es0177}. Besides exploration within condensed matter systems \cite{DiVincenzo1991quasicrystals,janot1994quasicrystals,Bindi2009natural,Jules2008archimedean,forster2013quasicrystalline,Zhiqing2018}, quasicrystalline order has since been observed across a wide range of materials. Prominent examples are QCs emerging in soft matter \cite{xianbing2004,
Hayashida2007,
talapin2009qc,
Tomonari1197QC}. 
Moreover, quasicrystalline order has been reported in various other contexts, such as quantum phase transitions \cite{Mivehvar2019}, vibrating granular materials \cite{plati2023quasicrystalline}, and has been recently discovered in defected regions (grain boundaries) between crystallites \cite{Devulapalli2024}. QCs are associated with unique properties, such as nonstick surfaces, low friction, and reduced thermal conductivity \cite{Macia2006,dubois2005useful,dubois2012properties}. Still, given the generality of this type of arrangement, the importance of modeling these systems extends far beyond these specific characteristics.

As with all systems that exhibit degrees of order, understanding the role of deformations and dislocations is crucial in QCs. Their unique lack of translational symmetry (periodicity) combined with the presence of rotational symmetry can be treated by extending classical elasticity theory. In brief, QCs can be described similarly to crystal upon considering a periodic hyperlattice in a higher-dimensional space from which the QCs can be obtained via projection \cite{Socoloar1986phonons,Bak86,guyot2001dislocations}. This also implies that dislocations and distortions are fully described within this hyperlattice framework, giving rise to two qualitatively distinct deformation fields: phonons and phasons. The former represents the conventional displacement field observed in crystals, while the latter is a unique concept in QCs, associated with local rearrangements of the quasicrystalline structure \cite{Socoloar1986phonons}. Both are crucial for understanding deformations and growth in QCs~\cite{guyot2001dislocations,Freedman2007,Kromer2012,yamada2016atomic,Galina2018,Zhang2024}.

Several approaches have been employed to model QCs. Continuum mechanics provided fundamental insights into the structural arrangement of aperiodic lattices and enabled a general assessment of their unique mechanical properties \cite{Levine85elasticity,Socoloar1986phonons,dubois2005useful,dubois2012properties}. At the same time, atomistic simulations have been instrumental in investigating atomic interactions and small-scale growth mechanisms (e.g., \cite{Dzugutov1993,Keys2007,deBoissieu2012,Engel2015,Schoberth2016}), as well as electronic properties (e.g., \cite{deLaissardiere1997,Brommer21022006,Mihalkovi2002,Kraj2005,Brommer2007}). 
Importantly, coarse-grained and mesoscale methods have also been developed to upscale microscopic information on quasicrystalline arrangements in a self-consistent and predictive manner.
These approaches include Monte Carlo simulations \cite{Widom1987,Kraj2005} and smooth field theories based on classical density functional theory and phase field crystal models \cite{Barkan2011,Rottler_2012,Archer2013,Achim2014,Barkan2014,xue2022atomic}.
Nevertheless, despite decades of research, several fundamental questions remain unanswered. 
For instance, even seemingly straightforward issues, such as dislocation formation at semi-coherent interfaces \cite{han2021formation}, continue to pose challenges and have not been fully understood. 
Recently, we proposed a mesoscale theoretical framework that realizes a coarse-graining of the microscopic density in QCs retaining the underlying quasicrystalline symmetry \cite{DeDonno2024}. This approach enables macroscopic modeling of QCs while inherently capturing dislocations, interfaces, and QC growth. It is built upon the main (shortest) reciprocal space vectors associated with QC symmetry and a minimal set of parameters, ensuring efficiency and generality as no information on dislocations is required \emph{a priori}. Proofs of concept have been delivered concerning the description of growth and elastic/plastic relaxation. In this work, we apply the framework to specifically model and investigate dislocations in QCs, analyzing their emergence and interactions mediated by the underlying microscopic symmetry. This leads to a novel and highly generalizable modeling approach for dislocations in systems exhibiting quasicrystalline order.

The manuscript is organized as follows. 
We first introduce the basics of the model and the description of QCs in Sect.~\ref{sec:introQCs} and \ref{sec:model}. 
We then study the emergence of dislocations at semicoherent interfaces forming between rotated inclusions and unrotated QCs domains, as discussed in Sect.~\ref{sec:rotation}. We show how the approach allows for the identification of dislocations by characterizing their Burgers vector, and we outline emerging fundamental differences with their crystalline counterparts. 
In Sect.~\ref{sec:strain}, we then focus on interfaces in strained systems. This allows us to investigate dislocation arrays and identify conditions for periodic arrays of dislocations in QCs. We use strained systems to discuss the energy of single dislocations, and the impact of phononic and phasonic deformations on the energetics of the dislocations. 
Dislocation stability is further discussed in Sect.~\ref{sec:dipoles}, where the interaction and motion of dislocation dipoles and quadrupoles are also analyzed. 
Our conclusions are summarized in Sect.~\ref{sec:conclusions}.

\section{Density Wave Representation of QCs and Deformations}
\label{sec:introQCs}

We consider the density-wave representation of QCs \cite{Levine85elasticity,Socoloar1986phonons}, i.e., their description as the result of the superposition of plane waves, set with zero average for simplicity:
\begin{equation}\label{eq:psir}
    \psi(\mathbf{r})=
    \sum_{j} \eta_j e^{\I \Gvn \cdot \rv} + \text{c.c.},
\end{equation}
with $\I$ denoting the imaginary unit, ``c.c." representing the complex conjugate, $\{\Gvn\}$ a discrete set of $N$ reciprocal-space vectors and $\mathbf{r} \in \Omega$ the coordinate in the space of definition of $\psi$. 
The complex amplitudes $\eta_{j}=\phi_{j}e^{\I\varphi_j}$ are slowly varying fields, with the real amplitude $\phi_j$ taking constant values in the bulk phases and varying at the interfaces between phases, and $\varphi_j$ accounting for deformations of $\psi$. 
An illustration of the density $\psi$ for a decagonal QC defined via the reciprocal space vectors as illustrated in Fig.~\ref{fig:model}(a) is shown in Fig.~\ref{fig:model}(b).

The description of the microscopic density via Eq.~\eqref{eq:psir} applies to both periodic crystals \cite{salvalaglio2022coarse} and QCs. 
The latter, however, are characterized by a discrete set of vectors $\{\Gvn\}$, which do not form a (reciprocal-space) lattice, unlike periodic crystals. 
In other words, the set of $\{\Gvn\}$ in Fig.~\ref{fig:model}(a) has more independent elements (four) than the dimension of the physical space (two). This results in characteristic aperiodic lattices and, importantly, prevents the application of constructions like Burgers vector circuits to determine topological charges associated with dislocations. QCs can still be described by defining a periodic lattice in a higher-dimensional space: the original space of definition of $\psi$, namely $\Omega\equiv \Omega^{\parall}$ with $\mathbf{r}\equiv \mathbf{r}^{\parall} \in \Omega^{\parall}$ called \emph{parallel} space, is complemented by the \emph{perpendicular} space $\Omega^{\perp}\ni \mathbf{r}^{\perp}$. The latter is constructed to obtain a periodic lattice in $\widetilde{\Omega}\equiv\Omega^{\parall} \cup \Omega^{\perp}$. 
Two deformation fields can then be defined, namely $\mathbf{u}\in \Omega^{\parall}$ and $\mathbf{w}\in \Omega^{\perp}$, called \emph{phonons} and \emph{phasons} respectively \cite{Socolar86unitcell}. 
With these definitions, the phase of complex amplitudes $\eta_n$ can be expressed as
\begin{equation}\label{eq:phases} 
    \varphi_j=\arg(\eta_j)=\Gvn^{\parall} \cdot \mathbf{u}+a\Gvn^{\perp} \cdot
    \mathbf{w}, 
\end{equation} 
with $\Gvn^{\parall}=\Gvn$, $\Gvn^{\perp}$ constructed such to obtain a periodic lattice in $\widetilde{\Omega}$ (see the case of decagonal QCs in \ref{fig:model}(a)), and $a$ a scaling factor depending on the symmetry of the QC.

The usefulness of the expression \eqref{eq:phases} and the underlying construction becomes evident when looking at perfect dislocations in QCs. 
They correspond to topological defects in the phase $\varphi_n$ with the topological charge given by 
\begin{equation}\label{eq:ointphase} 
2\pi s_j = -\oint d\varphi_j = 
-\oint \Gvn^{\parall} \cdot d\mathbf{u} -\oint a\Gvn^{\perp} \cdot d\mathbf{w}= 
-(\Gvn^{\parall} \cdot \bv^{\parall}+a\Gvn^{\perp} \cdot \bv^{\perp}), 
\end{equation}
with $s_j$ the (integer) winding number, and $\bv^{\parall}=\oint {\rm d}\mathbf{u}$ and $\bv^{\perp}=\oint {\rm d}\mathbf{w}$ Burgers vectors in $\Omega^{||}$ and $\Omega^{\perp}$, respectively. The minus sign follows the convention of the definition of the Burgers vector \cite{SKOGVOLL2022104932}.

Both phonons and phasons contribute to the elastic energy of QCs. For small distortions, the elastic energy density of QCs reduces to the quadratic form\footnote{we imply summation over repeated indices} 
\cite{DingPRB93},
\begin{equation}\label{eq:elast_form}
    e(\nabla\mathbf{u},\nabla\mathbf{w})=
    \frac{1}{2} C_{ik\ell m}\varepsilon_{ik}\varepsilon_{\ell m} +
    \frac{1}{2} K_{ik\ell m}\partial_k w_i\partial_m w_\ell +
    \frac{1}{2} R_{ik\ell m}\varepsilon_{ik}\partial_m w_\ell +
    \frac{1}{2} R_{ik\ell m}^\prime\partial_k w_i\varepsilon_{\ell m},
\end{equation}
with $\varepsilon_{\ell m}= \frac{1}{2}(\partial_\ell u_m + \partial_m u_\ell) \equiv \varepsilon_{\ell m}^\parallel$ and $\partial_m w_\ell\equiv \varepsilon_{\ell m}^\perp$ the phononic and phasonic strains, which can be used to explicit constitutive relations \begin{equation}\label{eq:stressANL}
\begin{split}    
    \sigma_{ik}^\parallel &= C_{ik\ell m} \varepsilon_{\ell m}^\parallel 
        +  R_{ik\ell m} \varepsilon_{\ell m}^{\perp},\\
    \sigma_{ik}^\perp   &= R_{ik\ell m}^\prime \varepsilon_{\ell m}^\parallel 
        +  K_{ik\ell m} \varepsilon_{\ell m}^{\perp}.   
\end{split}
\end{equation}

\section{Modeling by Amplitude Equations}
\label{sec:model}

The free energy for a set of $N$ amplitudes $\eta_j$ in Eq.~\eqref{eq:psir} encoding a specific crystalline or quasicrystalline symmetry can be derived through coarse-graining of the Swift-Hohenberg energy functional \cite{Swift1977,Elder2002}, and exploiting the fact that quasi unit cells featuring overlapping local motifs can be defined  \cite{Socolar86unitcell,steinhardt1998experimental}. 
It reads \cite{DeDonno2024} 
\begin{equation}\label{eq:free-energy}
\begin{split}    
    F_{\eta,P} &= \int_{\Omega} \bigg[\sum_{j} A|\mathcal{G}_{j}\eta_j|^2 + \sum_{p=2}^P  B_p\zeta_p\bigg]\ d\mathbf{r},
\end{split}
\end{equation}
with $A,B_p$ model parameters, $\mathcal{G}_{j}=\nabla^2+2{\rm \I}\Gvn\cdot\nabla$, and $\zeta_p$ polynomials defined as
\begin{equation}\label{eq:pol_resonance}
\zeta_p=\sum_{\boldsymbol{\alpha}_p} \eta_{\alpha_1}\eta_{\alpha_2}\ldots\eta_{\alpha_p}\delta_{\mathbf{0},\kv_{\alpha_1}+\kv_{\alpha_2}+\ldots+\kv_{\alpha_p}}, 
\end{equation}
with $\boldsymbol{\alpha}_p=\{\alpha_1,\alpha_2,\ldots,\alpha_p\}$ a multi-index with entries $\alpha_i$ taking values from $-N$ to $+N$ excluding zero (implying a labelling of $\eta_j$ and $\Gv_{j}$ by $j=1,\dots,N$), 
$\eta_{-j}=\eta_{j}^*$, and $\Gv_{-j}=-\Gv_{j}$. We remark that the quantity 
\begin{equation}\label{eq:orderPHI}
\zeta_2=\sum_{p,q=-N}^N\eta_p\eta_q \delta_{\mathbf{0},\kv_p+\kv_q}
    =2\sum_{j} |\eta_j|^2=\Phi,
\end{equation}    
directly describes the QC order; namely, it is constant in the crystalline bulk phases, decreases at defects and interfaces, and vanishes in disordered phases. 
The polynomial degree $P$ must be even to ensure that the free energy as a function of the amplitudes has a global minimum and, therefore, a stable phase.
$P=4$ is enough to describe a number of crystalline phases. 
QCs however require a higher polynomial degree, 
so that there is at least one non-trivial combination of reciprocal lattice vectors for which $\delta_{\mathbf{0},\kv_{\alpha_1}+\kv_{\alpha_2}+\ldots+\kv_{\alpha_p}} \neq 0 $ in Eq.~\eqref{eq:pol_resonance}.

From the definition of the free energy, Eq.~\eqref{eq:free-energy}, stress fields in both parallel and perpendicular spaces as introduced in Sect.~\ref{sec:introQCs} can be derived as $\sigma_{ik}^{\parall}=\delta F / \delta (\partial_k u_i)$ and $\sigma_{ik}^{\perp}=\delta F / \delta (\partial_k w_i)$ and read \cite{DeDonno2024}
\begin{equation}\label{eq:stressAPFC}
\begin{split}
    {\sigma_{ik}^{\rm \parall}} &= 
    8 \, \phi^2_0 A \sum_{j} \Gpar_{j,i}\Gpar_{j,k}\Gpar_{j,\ell}\partial_\ell\varphi_j, \\
    {\sigma_{ik}^{\rm \perp}} &= 
    8 \, \phi^2_0 A \sum_{j} \Gperp_{j,i}\Gpar_{j,k}\Gpar_{j,\ell}\partial_\ell\varphi_j,
\end{split}
\end{equation}
with $\phi_0$ the (real) value of amplitudes in the relaxed bulk \cite{salvalaglio2022coarse}, and $\mathbf{G}_{j,i}$ the $i$-th component of the $j$-th reciprocal space vector.
By expressing derivatives of the phases $\varphi_n$ in terms of $\varepsilon^{\parall}$ and  $\varepsilon^{\perp}$, the constitutive relation \eqref{eq:stressANL} are recovered with elastic constants
\begin{equation}\label{eq:elconstants}
\begin{split}
C_{ik\ell m} &=  8 A\phi_0^2\sum_j  \Gpar_{j,i}\Gpar_{j,k}\Gpar_{j,\ell}\Gpar_{j,m},\\
R_{ik\ell m} &= 16 A\phi_0^2\sum_j  \Gpar_{j,i}\Gpar_{j,k}\Gpar_{j,\ell}\Gperp_{j,m}, \\
K_{ik\ell m} &= 16 A\phi_0^2\sum_j  \Gpar_{j,i}\Gperp_{j,k}\Gpar_{j,\ell}\Gperp_{j,m}.
\end{split}
\end{equation}

The overdamped dynamics of amplitudes approximating the conservative dynamics of the microscopic density $\psi$ \cite{salvalaglio2022coarse} reads
\begin{equation}\label{eq:dynamicseta}
    \frac{\partial \eta_j}{\partial t}=
    -|\Gvn|^2\frac{\delta F_{\eta,P}}{\delta \eta_j^*}=
    -|\Gvn|^2 \bigg(A \mathcal{G}_j^2\eta_j + \sum_{p=2}^P B_p \partial_{\eta_n^*} \zeta_p\bigg).
\end{equation}

\subsection{Decagonal Quasicrystals}

\begin{figure}
    \centering
    \includegraphics[width=\textwidth]{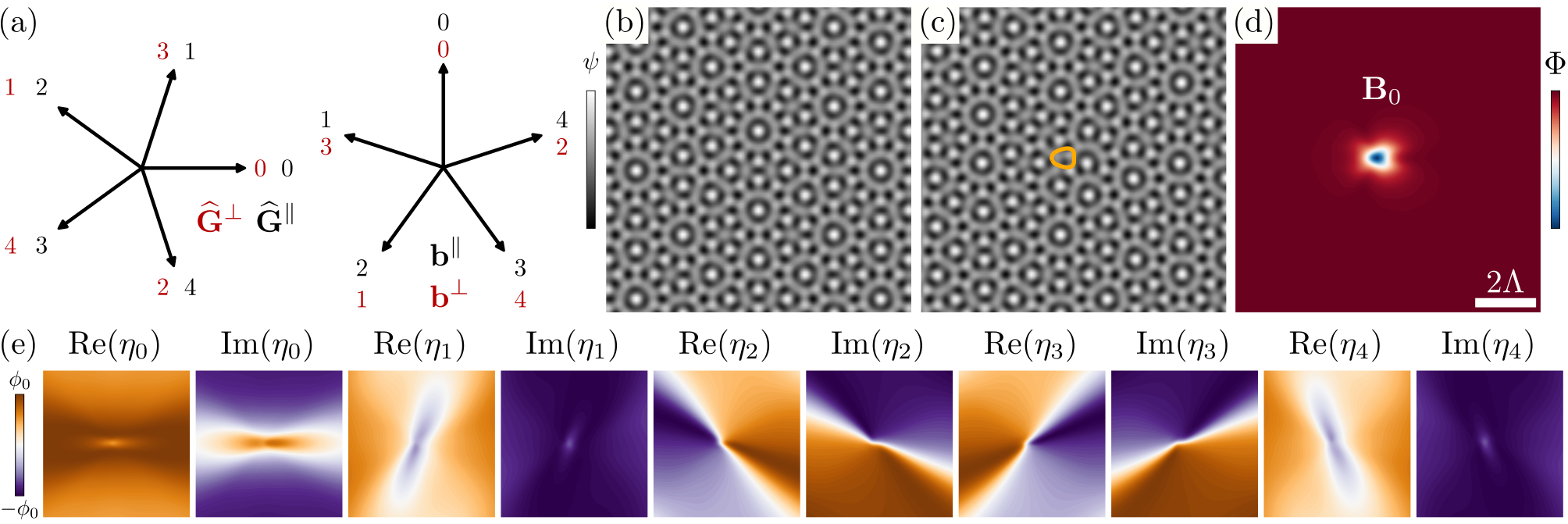}
    \caption{
    \textit{Modeling of Decagonal QCs by amplitude equations}. 
    (a) Reciprocal vectors and Burgers vectors in parallel and perpendicular space, as defined by Eq.~\eqref{eq:geo_vec_par} and Eq.~\eqref{eq:burgers}. 
    (b) Density field of a bulk decagonal QCs as described by Eq.~\eqref{eq:psir} with $\mathbf{G}_n\equiv\mathbf{G}_n^\parall$ from panel (a). (c) Density field of the decagonal QCs in (b) hosting a dislocation with Burgers vector $\mathbf B_0$ (see definitions in Sect.~\ref{sec:burgers}).
    The yellow line represents the position of the dislocation via a representative isoline of the field $\Phi = \sum_j |\eta_j|^2$.
    (d) Order parameter $\Phi$ in the same region as panel (c).
    (e) Real and imaginary parts of each amplitude in the same regions as panels (c, d). Note that the magnitude of $\eta_2$ and $\eta_3$ vanishes at the dislocation, as both real and imaginary parts are zero there. 
    Simulation parameters $A=1$, $B_2=0.02$, $B_4=0$, $B_5=-100$, $B_6=0.1$.
    }
    \label{fig:model}
\end{figure}

In this work, we study dislocations in decagonal QCs.  
The reciprocal lattice vectors encoding their microscopic symmetry are \cite{Levine85elasticity} 
\begin{equation}\label{eq:geo_vec_par}
    \begin{split}
        \Gvn^\parall &= \left[\cos(2\pi j/5),\sin(2\pi j/5)\right], \qquad j=0,...,4 \\
        \Gvn^\perp &= a\mathbf{G}_{3j}^\parall,
    \end{split}
\end{equation}
with $a=(1+\sqrt{5})/2$. The labeling from $0$ to $N-1$ is considered for consistency with previous literature on decagonal QCs (expressions in Eqs.~\eqref{eq:pol_resonance} and \eqref{eq:orderPHI} can be readily applied by shifting the index $j\rightarrow j+1$).
A modulo five operation is implied in all the indices such that $\mathbf{G}_{j+5i}^\parall = \Gvn^\parall \, \forall i \in \mathbb{Z}$, and the same for the other vectors. 
In Fig.~\ref{fig:model}(a) we show a plot of $\mathbf{G}_j^\parallel$ and $\mathbf{G}_j^\perp$. 
In Ref.~\cite{DeDonno2024}, we showed that averaging the microscopic density of a QC over a characteristic coarsening length $\Lambda$ produces a uniform field.
For the reciprocal lattice vectors in Eq.~\eqref{eq:geo_vec_par}, we get $\Lambda \sim 12.59$.  In what follows, we use $\Lambda$ as a unit length. 
We remark that the symmetry defined via the reciprocal lattice vectors is the only input of the amplitude framework related to the symmetry of QCs. 
Due to their ten-fold rotational symmetry, decagonal QCs can be described as stable phases of a free energy as in Eq.~\eqref{eq:free-energy} with $P=6$:
\begin{equation}\label{eq:free-energy5}  
    F_{\eta,6} = \int_{\Omega} \bigg[\sum_{j} A|\mathcal{G}_j\eta_j|^2 + 
B_2\zeta_2+B_3\zeta_3+B_4\zeta_4+B_5\zeta_5+B_6\zeta_6\bigg]\ d\mathbf{r},
\end{equation}
with polynomial terms following Eq.~\eqref{eq:pol_resonance} reading 
\begin{equation}\label{eq:supp:zeta_5_6}
    \begin{split}
    \zeta_2&\stackrel{\text{Eq.}~\eqref{eq:orderPHI}}{=}\Phi,
    \\
    \zeta_3& \equiv 0,
    \\
    \zeta_4&= 3 \Phi^2 -\frac{3}{2} \sum_j |\eta_j|^4, 
    \\
    \zeta_5^{\mathcal{D}} &= \bigg(\prod_j \eta_j\bigg)+\mathrm{c.c.}, 
    \\[1.5ex]
    \zeta_6^{\mathcal{D}} &= 720 \sum_{\substack{i\\j>i\\k>j}} |\eta_i|^2|\eta_j|^2|\eta_k|^2 + 180\sum_{\substack{i\\j\neq i}}|\eta_i|^4|\eta_j|^2
    +20\sum_i|\eta_i|^6. 
    \end{split}
\end{equation}

The simulation of a QC hosting a dislocation obtained by numerically solving Eq.~\eqref{eq:dynamicseta} is illustrated in Fig.~\ref{fig:model} (c-e). Such a numerical solution is obtained by a Fourier pseudo-spectral method enforcing periodic boundary conditions (see also \cite{salvalaglio2022coarse}).  
In this and all simulations reported below, the timestep is set to $\Delta t = 1$. The considered spectral method realizes a uniform spatial discretization, which is set to ensure ten mesh points per $\Lambda$.
The initial condition for the amplitudes is set by defining the amplitude phases $\varphi_n$ as in Eq.~\eqref{eq:phases}, with $\mathbf u$ and $\mathbf w$ (phonons and phasons) given by the elasticity solution for a dislocation dipole in an infinite QC as reported in Ref.~\cite{PialiPRB1987}. 
Then, Eq.~\eqref{eq:dynamicseta} is solved numerically to relax such an initial condition. 
Note that the amplitudes vary on significantly larger lengthscales than $\psi$, and $\psi$ can always be reconstructed from the amplitudes via Eq.~\eqref{eq:psir}.  
Fig.~\ref{fig:model}(c) shows such a reconstruction. 
Fig.~\ref{fig:model}(d) shows $\Phi$ computed from Eq.~\eqref{eq:orderPHI}, which decreases at the dislocation as the result of zeros of the magnitude of some of the amplitudes having singular phases. 
Single complex amplitudes are illustrated in Fig.~\ref{fig:model}(e) via both the real and imaginary parts. 
In particular, two amplitudes, namely $\eta_2$ and $\eta_3$, carry topological charges associated with the dislocation. This is manifested by the zero lines $\text{Re}(\eta)=0$ and $\text{Im}(\eta)=0$, corresponding to the branch cuts in the phase $\varphi = \arg(\eta)$. The dislocation is localized at the intersection of the zero lines. In general, zero lines correspond to codimension 1 manifolds, and dislocations to codimension 2 manifolds, such as planes and lines respectively in 3D. 
In a similar manner, complex amplitudes of sampling beams are often used to detect dislocations in crystals \cite{Phillips01062011,PHILLIPS20111483}. Note that, indeed, dislocations in ordered systems as (quasi-)crystals can be described as superposition of dislocations in the stripe phases corresponding to each Fourier mode in Eq.~\eqref{eq:psir} \cite{SkogvollNPJ2023}.

\subsection{Burgers vectors}
\label{sec:burgers}
For the analysis reported in the following, it is useful to define a 4-dimensional Burgers vector $\mathbf B_n$, which can be written as the direct sum of the two-dimensional Burger vectors in parallel and perpendicular space (see also Fig.~\ref{fig:model}(a)):
\begin{equation}\label{eq:burgers}
\begin{split}
    \mathbf B_n &= \mathbf{b}_n^\parallel \oplus \mathbf{b}_n^\perp ,
    \\
    \mathbf{b}_n^\parallel &= r_b(-\sin(2\pi n/5), \cos(2\pi n/5)),
    \\
    \mathbf{b}_n^\perp &= r_b(-\sin(6\pi n/5), \cos(6\pi n/5)),
    \\
\end{split}
\end{equation}
where $n = 0,...,4$ and $r_b = \frac{8}{5}\pi \sin(4\pi/5)$.
By construction, the following identities hold:
\begin{equation}\label{eq:B-identities}
\begin{split}
    \widehat{\mathbf{b}}^\parallel_n &
        = \widehat{\mathbf{b}}^\parallel_{n+1} + \widehat{\mathbf{b}}^\parallel_{n-1} 
        = -(\widehat{\mathbf{b}}^\parallel_{n+2} + \widehat{\mathbf{b}}^\parallel_{n-2}),
    \\
    \widehat{\mathbf{b}}^\perp_n &
        = \widehat{\mathbf{b}}^\perp_{n+2} + \widehat{\mathbf{b}}^\perp_{n-2} 
        = -(\widehat{\mathbf{b}}^\perp_{n+1} + \widehat{\mathbf{b}}^\perp_{n-1}),
\end{split}
\end{equation}
%
where $\widehat{\mathbf{v}} = \mathbf{v}/{||\mathbf{v}||}$.
Eq.~\eqref{eq:burgers} describes the shortest Burgers vectors allowed. 
In analogy with dislocations in crystals, these are expected to correspond to the dislocations with the lowest energy. When considering composite Burgers vectors, it is helpful to distinguish between compositions of the type $\mathbf{B}_n + \mathbf{B}_{n+1}$, and $\mathbf{B}_n + \mathbf{B}_{n+2}$. The length of the four-dimensional Burgers vector is the same for both types. However, the length of the Burgers vectors in the parallel and perpendicular subspaces is different: 
\begin{equation}\label{eq:B-len}
\begin{split}
    || \mathbf{b}^\parallel_n + \mathbf{b}^\parallel_{n+1} || &> 
    || \mathbf{b}^\perp_n + \mathbf{b}^\perp_{n+1} ||, \\
    || \mathbf{b}^\parallel_n + \mathbf{b}^\parallel_{n+2} || &< 
    || \mathbf{b}^\perp_n + \mathbf{b}^\perp_{n+2} ||.    
\end{split}
\end{equation}
%
In Section~\ref{sec:strain}, we show that this difference significantly affects the energetics of the dislocations.
Finally, we remark that the distinction between $\mathbf{B}_n + \mathbf{B}_{n+1}$, and $\mathbf{B}_n + \mathbf{B}_{n+2}$ also exhausts all possible combinations of three unique Burgers vectors, since, by construction, $\mathbf{B}_n + \mathbf{B}_{n+1} = -(\mathbf{B}_{n+2} + \mathbf{B}_{n+3} + \mathbf{B}_{n+4})$. 
Using the same geometrical argument, the sum of four unique Burgers vector yields a primitive Burgers vector: $ \mathbf{B}_{n+1} + \mathbf{B}_{n+2} + \mathbf{B}_{n+3} + \mathbf{B}_{n+4} = - \mathbf{B}_n$. 

\subsection{Tracking dislocations via amplitudes}
\label{sec:tracking}

Describing ordered systems via complex amplitudes allows for an advanced analysis of dislocations. 
On the one hand, the field $\Phi$ from Eq.~\eqref{eq:orderPHI} allows for detecting dislocations straightforwardly. A demonstration is given in Fig.~\ref{fig:model}: a dislocation in a quasicrystal, when resolved at microscopic scales, is typically challenging to detect; see Fig.~\ref{fig:model}(c). However, it can be easily identified using the field $\Phi$, as illustrated by the yellow curve, representing the isoline $\Phi/\Phi_{\mathrm{MAX}} = 0.7$, and fully reported in Fig.~\ref{fig:model}(d). 
On the other hand, information on the topological charge of dislocations can be extracted \cite{SKOGVOLL2022104932,SkogvollNPJ2023}. In particular, one can determine the Burgers vector densities $\mathcal{B}^{{\perp}}$ and $\mathcal{B}^{{\parall}}$ for a dislocation at $\rv_0$ via the zeros of $\eta_n$ corresponding to singularities in the phases $\varphi_n$. 
See, for instance, that in Fig.~\ref{fig:model}(e) amplitudes $\eta_2$ and $\eta_3$ vanish at the dislocation core, pointing at singular $\varphi_2$ and $\varphi_3$. 
The system of equations obtained with Eq.~\eqref{eq:burgers} for both $n=2$ and $n=3$ can be then solved for the Burgers vectors, which indeed result into $\mathbf{B}_0$, as imposed in the initial condition in this case, but not known \emph{a priori} for dislocations that form spontaneously.

The Burgers vector densities $\mathcal{B}^{{\parall}}$ and $\mathcal{B}^{{\perp}}$ in parallel and perpendicular space, respectively, are represented by Dirac-delta functions centered at the dislocation position $\mathbf r_0$. By a coordinate transformation from $\rv$ to [Re($\eta_j$),Im($\eta_j$)], we can formulate an equivalent representation of the Burgers vector densities located at the zeros of the complex amplitudes and with values determined by the $D$-fields, namely
\begin{equation}\label{eq:dislodensity}
\begin{split}
    \mathcal{B}^{||}&=\bv^{{\rm ||}}\delta (\rv-\rv_0)= 
    -\frac{4\pi}{N|\bv^{||}|^2}\sum_j \Gvn^\parallel \mathrm D_j \delta(\eta_j), 
    \\
    \mathcal{B}^{\perp}&=\bv^{{\rm \perp}}\delta (\rv-\rv_0)= 
    -\frac{4\pi}{Na^2|\bv^{\perp}|^2}\sum_j \Gvn^\perp \mathrm D_j \delta(\eta_j),
\end{split}
\end{equation}
as obtained by contracting Eq.~\eqref{eq:ointphase} with $\Gvn^\parallel$ and $\Gvn^\perp$, and
\begin{equation}\label{eq:Dfield}
\mathrm D_j = \frac{\epsilon_{ik}}{2\I}\partial_i\eta_j^* \partial_k \eta_j, 
\end{equation} 
with $\epsilon_{ik}$ the 2D Levi-Civita symbol.
The fields D$_j$ are formally the determinant of the coordinate transformation from $\rv$ to [Re($\eta_j$), Im($\eta_j$)]. In practice, ${\rm D}_j$ is a smooth field that captures the spatial changes in $\eta_j$ in the presence of a topological charge. Namely, ${\rm D}_j\ne 0$ at the intersection of the zero lines in $\eta_j$ (i.e. $\text{Re}(\eta_j) =0$ and $\text{Im}(\eta_j) =0$) as shown in Fig.~\ref{fig:model}(e).   

Eqs.~\eqref{eq:Dfield} and \eqref{eq:dislodensity} then allow us to reconstruct which dislocations are encoded in a given set of complex amplitudes $\eta_j$. In what follows, we exploit the D fields to characterize emerging dislocations and track their dynamics.

\section{Rotated inclusions} \label{sec:rotation}

\begin{figure}
    \centering
    \includegraphics[width=\textwidth]{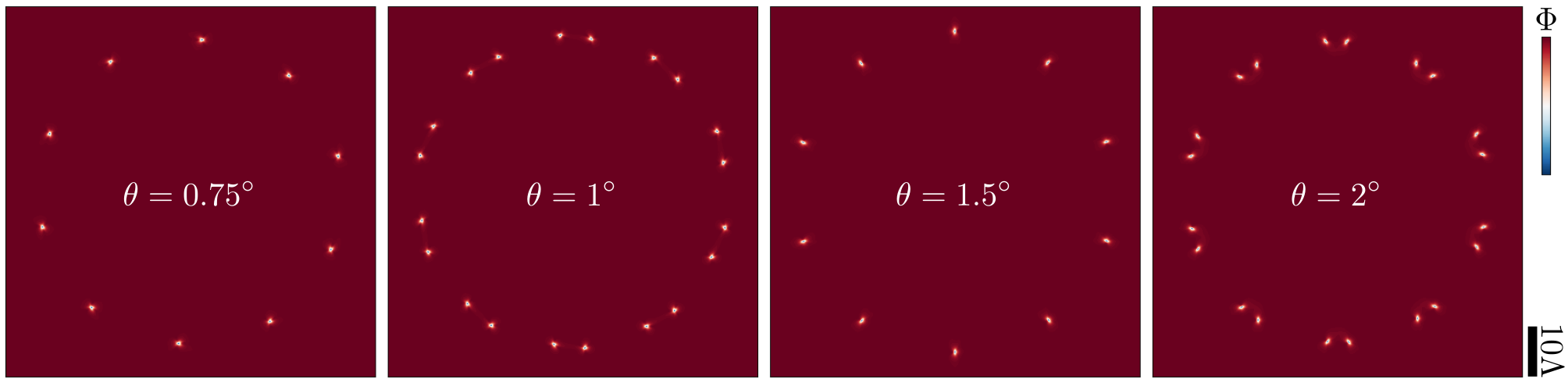}
    \caption{
    \textit{Circular inclusion.}
    Plots of the order parameter $\Phi$ for a circular inclusion of diameter 60 $\Lambda$ rotated by a small angle $\theta$ with respect to the surrounding matrix.
    Dislocations are characterized by localized minima in $\Phi$.
    The inclusion shrinks as the system evolves. Snapshots are taken $10^4$ time units into the evolution. 
    Simulation parameters are the same as in Fig.~\ref{fig:model}.
    }
    \label{fig:inclusion}
\end{figure}

In a 2D crystal, a rotation by a (tilt) angle $\theta$ can be described by complex amplitudes given by
\begin{equation}\label{eq:rota}
\begin{split}
    \eta_j &= \phi_0 e^{-\ii \delta\mathbf{G}_j \cdot \mathbf{r}},\\
    \delta\mathbf{G}_j &= \mathcal{R}_\theta \cdot \mathbf{G}_j  - \mathbf{G}_j,
\end{split}
\end{equation}
with $\mathcal{R}_\theta$ the counterclockwise rotation matrix by an angle $\theta$. 
Indeed, inserting this expression into Eq.~\eqref{eq:psir} yields a density wave representation with rotated reciprocal space vectors $\mathcal{R}_\theta \cdot \Gvn$. 
In Ref. \cite{DeDonno2024}, while studying the growth of quasicrystalline seeds in a melt, we have shown that Eq.~\eqref{eq:rota} applies straightforwardly to quasicrystals. In other words, when imposing a rotation in a quasicrystal, one should only rotate the reciprocal lattice vectors in parallel space.
In this section, we study the dislocations that appear at the boundary of a circular inclusion in a quasicrystalline matrix. 
For each dislocation, we can determine its Burgers vector via the D fields introduced in Eq.~\eqref{eq:Dfield}.
The inclusion has a diameter of 60 $\Lambda$, and it is rotated with respect to the surrounding matrix by a small angle $\theta$. 
The rotation is obtained from Eq.~\eqref{eq:rota}, by imposing $\theta\neq 0$ in the inclusion, and $\theta=0$ in the matrix.
In the early stages of the dynamics, the system creates a number of dislocations at the boundary of the inclusion to accommodate the mismatch between the rotated and unrotated regions. 
The system evolves by reducing the surface of the boundary; therefore, the inclusion shrinks and eventually disappears.  

In Fig. \ref{fig:inclusion}, we show plots of the order parameter for different mismatch angles. 
Each snapshot is taken $10^4$ time units into the evolution. 
We note that in all cases the dislocations form ten radially symmetric clusters, which can be ascribed to the tenfold rotational symmetry of the considered QC. 
The composition of each cluster, however, depends on the mismatch angle. 
For small mismatch ($\theta=0.75^\circ$), each cluster contains a single low-energy dislocation, with the shortest Burgers vectors $\mathbf B_n$. 
As the mismatch increases ($\theta=1^\circ$), a second low-energy dislocation appears in each cluster. 
Increasing the mismatch further ($\theta=1.5^\circ$) results in a single dislocation appearing per cluster. 
The drop in the total number of dislocations as the relative rotation increases can only be explained by an increase of the Burgers vector length associated with each dislocation. 
Indeed, the dislocation analysis via the D field (see Sect.~\ref{sec:tracking}) reveals that these are composite dislocations, with Burgers vectors of the type $\mathbf B_n + \mathbf B_{n+1}$.
Increasing the mismatch again ($\theta=2^\circ$) results in one more composite dislocation appearing per cluster.

These results point to a peculiar behavior concerning the plastic deformation of QCs. 
For small tilt angles, at the onset of plastic deformation (insertion of dislocations), the system resolves the mismatch between inclusion and matrix via the lowest-energy dislocations. This is a usual behavior for conventional crystalline materials, too. In these systems, further increasing the relative rotation leads to the nucleation of additional dislocations of the same type, up until the limit of large-angle rotation, where the spacing between dislocations is very small, and the minimum energy configuration consists of different local rearrangements, namely large-angle grain boundaries \cite{HAN2017186}.
However, our simulations show that in QCs a different hierarchy of dislocations is obtained. 
There exists a mismatch threshold over which the system favors the nucleation of a smaller number of high-energy dislocations instead of a large number of low-energy dislocations. Different dislocation configurations may form and coexist in equilibrium and metastable states. 
We also remark that the considered framework realizes relaxation of initial conditions and dynamics following a weighted gradient flow scheme, thus ensuring energy decrease over solution trajectories. Moreover, this observation aligns with an earlier finding based solely on geometric considerations, which predicted the existence of at least two distinct types of dislocations in small-angle grain boundaries in QCs and, therefore, highlighted the need for a generalized framework extending the Read–Shockley model for small-angle grain boundary energy as a function of tilt angle \cite{QCsmallGB_PhysRevLett.62.2699}.  
This behavior is unique to QCs and is inherently linked to their extended deformation mechanisms and associated energetics. The following section explores the energy of isolated dislocations of different types, focusing on those that form at interfaces between strained QCs.

\section{Strained systems}
\label{sec:strain}

\begin{figure}
    \centering
    \includegraphics[width=\textwidth]{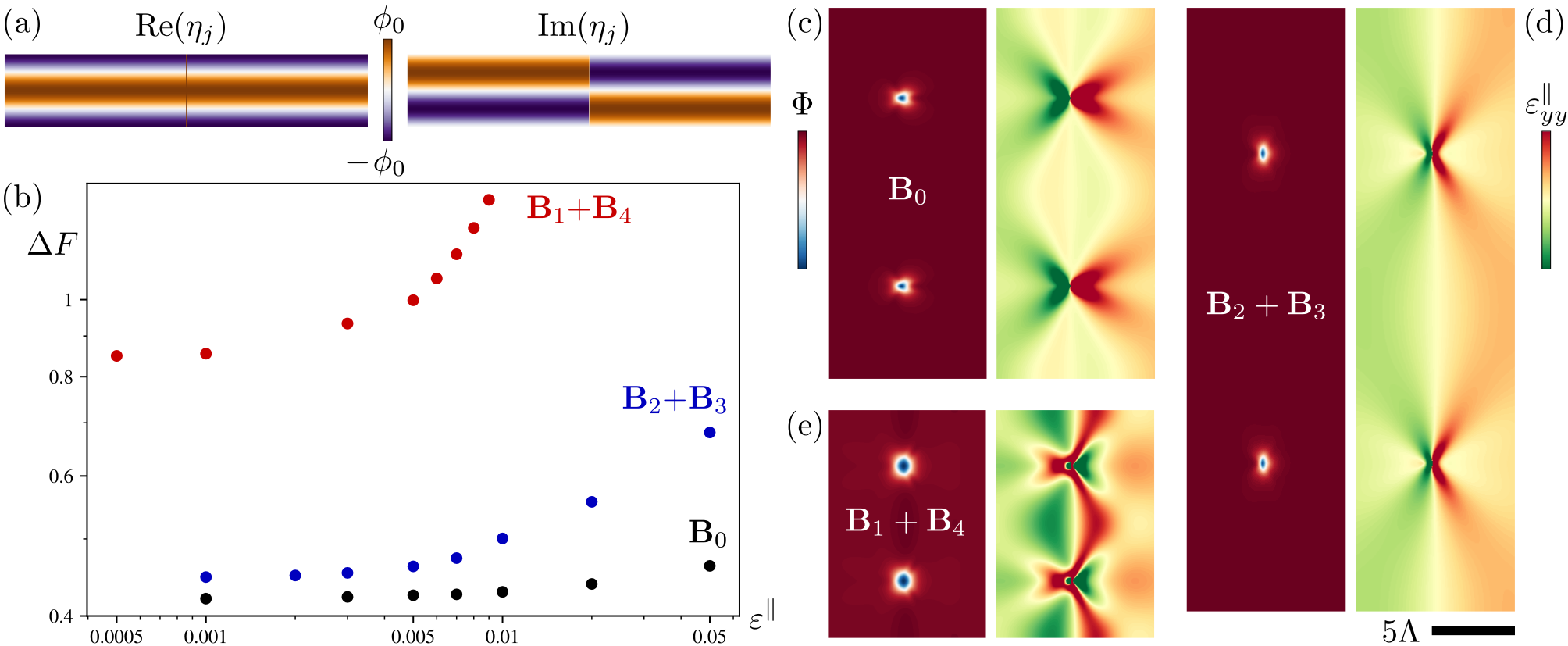}
    \caption{
    \textit{Strained systems.}
    (a) Initial condition: plot of real and imaginary parts of an oscillating amplitude. 
    $\phi_0$ is the value of the amplitude in the relaxed bulk.
    Amplitudes oscillate once along $y$ and are constant along $x$.
    (b) Log-log plot of the dislocation energy against the strain for different types of dislocations featuring Burgers vector as reported in the plot.
    The dislocation energy is calculated as the difference between the free energy of the system hosting four dislocations and the free energy of a relaxed bulk system with the same size.
    The energy curves saturate for small strains as the interaction energy between dislocations becomes negligible.
    (c,d,e) Plots of the order parameter $\Phi$ (left) and strain component $\epar_{yy}$ (right) for dislocations with different Burgers vectors (considered in panel (b), too).
    Simulation parameters are the same as in Fig.~\ref{fig:model}.
    }
    \label{fig:panino}
\end{figure}

Dislocations form at interfaces between ordered arrangements having a relative mismatch due to different lattice spacing or strain states. Similarly to the description of rotations, such deformations can be encoded in varying complex amplitudes. For a uniform volumetric strain $\varepsilon_{\ell m}^{\parall,\perp}= \bar{\varepsilon}^{\parall,\perp}\delta_{\ell m}$, the amplitudes are $\eta_j = \phi_0 e^{-\ii \boldsymbol{\nu} \cdot \mathbf{r}}$ with 
\begin{equation}\label{eq:strainQC}
    \boldsymbol{\nu}_j = \bar{\varepsilon}^{\parall}\mathbf{G}^\parallel_j  + a \bar{\varepsilon}^{\perp}\mathbf{G}^\perp_j . 
\end{equation}
For now, we consider $\epar,\,\eperp$ as independent parameters.
Evaluating explicitly the expression above for a uniaxial strain along $x$, for a QCs described in the relaxed state by reciprocal space vectors as in Fig.~\ref{fig:model}(a), yields:
\begin{equation}
\begin{split}
    \nu_0^x &= \epar + a \eperp,\\
    \nu_1^x = \nu_4^x &= \frac{1}{2} \left((a-1)\epar-\left(a+1\right) \eperp\right),\\
    \nu_2^x = \nu_3^x &= \frac{1}{2} \left( \eperp -a\,\epar\right).
\end{split}
\end{equation}
Doing the same for a uniaxial strain along $y$ yields:
\begin{equation}\label{eq:strainY}
\begin{split}
    \nu_0^y &= 0, \\
    \nu_1^y = \nu_4^y &= \frac{1}{2} \sqrt{\frac{1}{2} \left(\sqrt{5}+5\right)} (\epar -\eperp),\\
    \nu_2^y = \nu_3^y &= \frac{1}{4} \left(\sqrt{10-2 \sqrt{5}} \epar +2 \sqrt{2 \sqrt{5}+5} \eperp\right),\\
\end{split}
\end{equation}

In what follows, we focus on the case of a unidirectional strain applied only along $y$. Then, from Eq.~\eqref{eq:strainY}, we can distinguish three special cases. 
If we choose $\epar,\,\eperp$ so that $\nu_1^y = \nu_4^y = 0$, the only oscillating amplitudes are $\eta_2$ and $\eta_3$. Conversely, if we choose $\epar,\,\eperp$ so that $\nu_2^y = \nu_3^y = 0$, the only oscillating amplitudes are $\eta_1$ and $\eta_4$. Finally, we can also choose $\epar,\,\eperp$ so that all amplitudes (excluding $\eta_0$) have the same $\nu$. 
The corresponding values of $\epar,\,\eperp$ are:
\begin{equation}\label{eq:eps_combo}
\begin{split}
    \eperp = \epar \quad &\to \quad \nu_1^y = \nu_4^y = 0, \\
    \eperp = -(a+1)\epar \quad &\to \quad \nu_2^y = \nu_3^y = 0, \\
    \eperp = (a-2)\epar \quad &\to \quad \nu_1^y = \nu_2^y = \nu_3^y = \nu_4^y. 
\end{split}
\end{equation}
We recall that the topological charge is carried by singularities in the phase of amplitudes; see Eq.~\eqref{eq:ointphase}. 
These may be initialized by imposing varying amplitudes with discontinuities. In the case discussed here, this corresponds to imposing oscillations in amplitudes with the frequencies reported above and abrupt changes of the phases of such oscillations. By controlling which amplitudes oscillate, we are then able to predict the Burgers vector of the dislocations that form in the system:
\begin{equation}
\begin{split}
    \nu_1^y = \nu_4^y = 0 \quad &\to \quad \mathbf B _0, \\
    \nu_2^y = \nu_3^y = 0 \quad &\to \quad \mathbf B _2+\mathbf B _3, \\
    \nu_1^y = \nu_2^y = \nu_3^y = \nu_4^y  \quad &\to \quad \mathbf B _1+\mathbf B _4. 
\end{split}
\end{equation}
Furthermore, when all oscillating amplitudes have the same frequency $\nu$, the system is periodic in space, with the period of the oscillation corresponding to a wavelength $\lambda = 2\pi / \nu$.
This implies the possibility of initializing periodic arrays of dislocations with a known Burgers vector.
Each oscillation period produces a pair of dislocations, at a distance $\lambda /2$ from one another. This distance can be controlled by varying $\epar$. As expected, $\lambda \sim 1/\epar$, meaning that a system with a small applied strain produces dislocations that are at a large distance from one another. 
The analysis reported in this section is of two-fold importance: On the one hand, deformations exist that lead to periodic arrays of dislocations (of one type) despite the underlying quasicrystalline (aperiodic) order. 
On the other hand, this deformation can be exploited to study the properties of one type of dislocation at the time. 

\subsection{Dislocation energies}
\label{sec:dislo}
As static systems hosting a low number of dislocations of the same type can be realized, we can exploit them to evaluate the energy of a single dislocation, which is here inspected via numerical simulations. 
Fig.~\ref{fig:panino}(a) shows the initial condition for an interface between mismatched QCs, with an interface normal oriented along the $x$-axis, and an initialized oscillating amplitude (both real and imaginary parts are shown) in a rectangular domain $[-L_x/2,L_x/2]\times[-L_y/2,L_y/2]$. The system size along $y$ corresponds to exactly one oscillation period, $L_y=2\pi/\nu^y_j$. The value can be obtained by inserting the appropriate line (depending on the investigated dislocation) of Eq.~\eqref{eq:eps_combo} into Eq.~\eqref{eq:strainY} to determine the oscillating frequency $\nu^y_n$. 
The discontinuity at the interface located at the center ($x=0$) in Fig.~\ref{fig:panino}(a) is obtained by setting $(\epar/2,\eperp/2)$ for $x<0$ and $-(\epar/2,\eperp/2)$ for $x>0$. Note that the oscillating frequency is the same in both cases, but a phase shift of $\pi$ is realized. This way, two dislocations form in the positions $(0,\pm L_y/4)$.
Periodic boundary conditions are applied to all boundaries so that two more dislocations with opposite Burgers vectors form at the (periodic) boundary with normal along the $x$-axis and at $\pm L_y/4$.
All amplitudes are constant along $x$ by construction, as non-zero strain is set only for $\varepsilon_{yy}^{\parall}$ and $\varepsilon_{yy}^{\perp}$, so we only need to ensure that the system is large enough to prevent dislocations from interacting too strongly. In all cases, we choose the horizontal size $L_x$ such that $L_x/L_y =5$.
Figure \ref{fig:panino}(c-e) shows plots of the order parameter $\Phi$ and the strain component $\epar_{yy}$ (obtained inverting and deriving Eq.~\eqref{eq:phases}; see also Ref.~\cite{DeDonno2024}) for pairs of dislocations with different Burgers vectors. 
In all cases, we take $\epar = 0.01$, and use Eq.~\eqref{eq:eps_combo} to determine $\eperp$ and Eq.~\eqref{eq:strainY} to determine $L_y$. 
The full extension of the system along the $y$-axis is shown, 
focusing along the $x$-axis on the region around the interface.  
Since the variation of $\eperp$ with respect to $\epar$ changes with the type of dislocation,
we remark that the same value of $\epar$ produces dislocations separated by different distances.

Fig.~\ref{fig:panino}(b) shows the dependence of the dislocation energy on the strain for different types of dislocations. 
The dislocation energy $\Delta F$ is calculated as the difference between the free energy of the system hosting four dislocations and the free energy of a relaxed bulk system of the same size. Thus, it comprises two contributions:
\begin{equation}
    \Delta F=N E_{\rm dislo} + E_{\rm interaction},
\end{equation}
with $E_{\rm dislo}$ the total energy (including core and elastic contributions) associated with the isolated dislocations, $N$ the number of dislocations (here, $N=4$), and $E_{\rm interaction}$ the elastic interaction due to the superposition of the elastic fields. 
Therefore, these curves tend to the energy of four isolated dislocations for $\epar \rightarrow 0$.
By comparing the energy values in Fig.~\ref{fig:panino}(b) for dislocations of different types, we can determine that the dislocation with Burgers vector $\mathbf{B}_1+\mathbf{B}_4$ has an energy which is roughly double that of a simple dislocation with Burgers vector $\mathbf{B}_0$. 
At the same time, the dislocation with Burgers vector $\mathbf{B}_2+\mathbf{B}_3$ is significantly closer in energy to the lowest energy one. 
This is remarkable, as the length of the four-dimensional Burgers vectors $\mathbf{B}_2+\mathbf{B}_3$ and $\mathbf{B}_1+\mathbf{B}_4$ is the same. 
However, from Eq.~\eqref{eq:B-len}, we know that the length of the Burgers vectors in the subspaces is different. 
Thus, we observe that the formation of a dislocation with Burgers vector with a shorter component in perpendicular space than in parallel space (such as $\mathbf{B}_2+\mathbf{B}_3$) is energetically favored with respect to a dislocation with $||\mathbf b^\parallel||<||\mathbf b^\perp||$, such as $\mathbf{B}_1+\mathbf{B}_4$.
In other words, the length of the projection of the Burgers vector in perpendicular space has a greater impact on the energetics of the dislocation than that in parallel space.
Still, the length of the four-dimensional Burgers vectors plays a role: $\mathbf B_0$ is shorter than both $\mathbf{B}_2+\mathbf{B}_3$ and $\mathbf{B}_1+\mathbf{B}_4$, and the corresponding dislocation has the lowest energy, consistently with the evidence discussed in Sect.~\ref{sec:rotation}.

\section{Dislocation dipoles}
\label{sec:dipoles}

\begin{figure}[t]
    \centering
    \includegraphics[width=\textwidth]{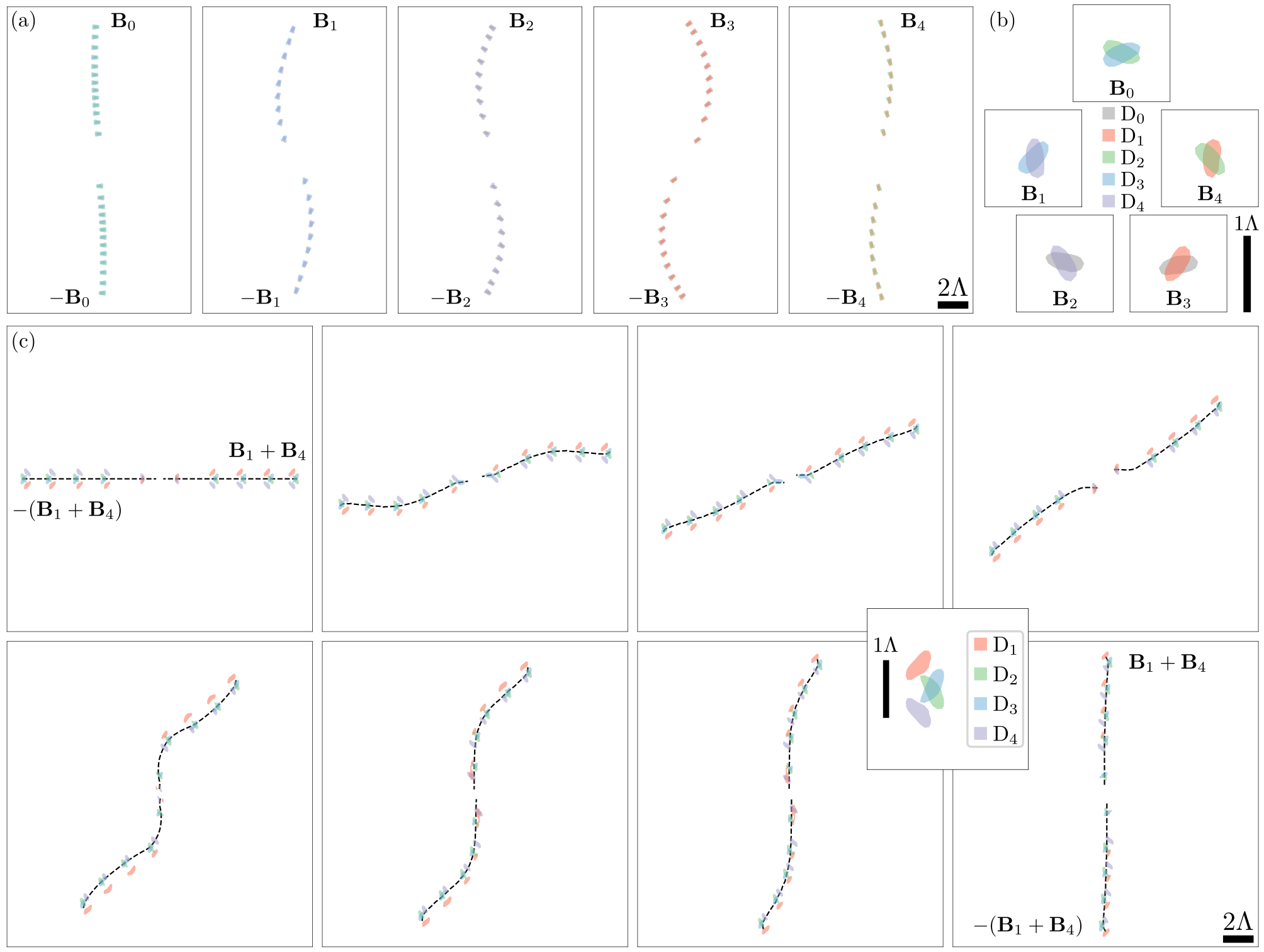}
    \caption{
    \textit{Dislocation dipoles.}
    (a) Plots of the D field for dislocation dipoles with Burgers vectors $\mathbf{B}_n$ over time.  
    Dislocations are initialized in the positions $(0,\pm 9\Lambda)$.
    Snapshots of the evolution are taken at intervals of $5\cdot 10^3$ time units. 
    (b) Arrangement of the D fields, Eq.~\eqref{eq:Dfield}, of isolated dislocations with the shortest Burgers vectors $\mathbf{B}_n$. 
    (c) Plots of the D field for a dislocation dipole with Burgers vector $\pm (\mathbf{B}_1 +\mathbf{B}_4)$. The dislocation trajectory is traced with a dashed line.
    The dislocations are initially positioned at opposite points of a circle with a diameter of $18\Lambda$.
    Snapshots of the evolution are taken at intervals of $10^4$ time units. 
    A detail of the arrangement of the D fields for an isolated dislocation is shown in the inset.
    Simulation parameters are the same as in Fig.~\ref{fig:model}.
    }
    \label{fig:dipole_type2}
\end{figure}

The systems studied in Sect.~\ref{sec:dislo} are stationary. To further inspect the stability of the dislocations observed therein, we consider dynamical settings involving dislocation dipoles. 
We recall that our amplitude model delivers a Peach-Koehler-type equation of motion for dislocations. The dislocation velocity was derived in Ref.~\cite{DeDonno2024}, assuming it is the same for dislocations in the parallel and perpendicular space \cite{Lubensky86dislomotion}.
Following the definitions used in this work, the $i$-th component of the velocity of a dislocation in a decagonal QC is 
\begin{equation}\label{eq:velocity}
        v_i = \frac{5A}{\pi} \epsilon_{ij}\left( \sigma^{\parall}_{jk}b_k^{\parall}+a^2\sigma^{\perp}_{jk}b_k^{\perp} \right),
\end{equation}
with $\boldsymbol{\sigma}^{\parall}$ and $\boldsymbol{\sigma}^{\perp}$ the phononic and phasonic stress reported in Eq.~\eqref{eq:stressAPFC}, consistent with previous studies on dislocation dynamics in QCs \cite{Lubensky86dislomotion,Agiasofitou_2010}.

The study we address in this section builds upon a few of the previous findings. 
First, dislocations with Burgers vector $\mathbf{B}_n$ and $\mathbf{B}_n+\mathbf{B}_{n+1}$ spontaneously emerge at interfaces between rotated QCs (see Sect.~\ref{sec:rotation}), whereas $\mathbf{B}_n+\mathbf{B}_{n+2}$ do not. 
Further, all these dislocations can form under suitable deformations or constraints and remain stable upon relaxation (see Sect.~\ref{sec:dislo}). 
Finally, the derivation of Eq.~\eqref{eq:velocity} holds rigorously for dislocations with Burgers vectors $\mathbf{B}_n$ and $\mathbf{B}_n+\mathbf{B}_{n+1}$, as they are expected to have the same velocity in both parallel and perpendicular space by construction (see Ref.~\cite{DeDonno2024} for a detailed discussion). However, this condition does not necessarily apply to dislocations with Burgers vector $\mathbf{B}_n+\mathbf{B}_{n+2}$.  
It thus remains an open question whether these dislocations remain stable, evolving similarly to the others, or instead dissociate into different dislocations once the constraints imposed in Sect.~\ref{sec:dislo} are removed.
Here, we inspect dislocations with different Burgers vectors, initialized via the solutions for $\textbf{u}$ and $\textbf{w}$ reported in Ref.~\cite{PialiPRB1987}, properly superposed to account for the two dislocations and then inserted in the equation for the amplitude phases, Eq.~\eqref{eq:phases}. 

In Fig.~\ref{fig:dipole_type2}(a), we report the evolution of dislocation dipoles with Burgers vectors $\pm \mathbf{B}_n$ by showing snapshots of the D fields equally spaced in time.
The initial position of the dislocations is $(0,\pm9\Lambda)$.
We remark that the trajectory varies with the orientation of the Burgers vector. 
In particular, the dislocations with Burgers vector $\pm\mathbf{B}_0$ move in a straight line, along the direction of the Burgers vector. This is akin to the glide motion in conventional crystals.
In Fig.~\ref{fig:dipole_type2}(b), we show details of the charge distribution for isolated dislocations. Colored regions are defined as the intervals in which $|\mathrm{D}_n| \geq 0.1 \mathrm{max}(\mathrm{D}_n)$. 
We note that each dislocation is composed of two non-zero D fields. 
Further, the two non-zero D fields mostly overlap, meaning that the topological charge is well localized in space.  
In Fig.~\ref{fig:dipole_type2}(c), we show the evolution of dislocation dipoles with Burgers vector $\pm(\mathbf{B}_1+\mathbf{B}_4)$. From Eq.~\eqref{eq:B-identities}, we recall that the Burgers vectors we are considering are parallel to $\mathbf{B}_0$ in parallel space and antiparallel to it in perpendicular space.
The initial position of the dislocations is $\pm 9 \Lambda (\cos(m\pi/16),\sin(m\pi/16)$, with $ m = 0,...,8$, corresponding to placing the dislocations opposite to one another on a circumference with diameter $18\Lambda$. 
We plot the dislocation trajectory with a dashed line and show snapshots of the D fields equally spaced in time. 
In all cases, the dislocations attract and eventually annihilate; however, the trajectory of the dislocations depends on the initial position. This behavior closely resembles the dynamic of dislocations in crystals, with the first and the last panel mimicking the motion by climbing and gliding (i.e., motion perpendicular or parallel to the Burgers vector), while all other cases involving a mixed-type motion between these two \cite{SalvalaglioJMPS2020}. 
We remark that the mesoscale description encoded in amplitude equations neglects a few effects, such as the activation barrier for climbing, while retaining a slower velocity for climbing than gliding,  thus mimicking high-temperature behaviors. 

The inset in Fig.~\ref{fig:dipole_type2}(c) shows the spatial arrangement of the (smooth) charges associated with amplitudes with singular phases. 
Notably, for the dislocation in question $\mathrm{D}_0\sim0$ everywhere in space. 
We observe that the dislocation is constituted by four regions associated with non-zero topological charges, which not only cover an extended area in space but are also delocalized. In other words, the zeros of the amplitudes remain bounded but are located in different positions in space, significantly differing from the case in Fig.~\ref{fig:dipole_type2}(a), as well as dislocations in crystals. 
As shown in Fig.~\ref{fig:dipole_type2}, the arrangement of the charges remains very similar during the whole evolution, provided that the dislocations are sufficiently far away from one another. This indicates that the charges associated with each amplitude move with the same velocity. Therefore, we find that these dislocations evolve as single objects and that a dynamic resembling the one dictated by Eq.~\eqref{eq:velocity} is realized, although for a configuration of bounded but not overlapping charges.
In Fig.~\ref{fig:dipole_type2}(c), we also show that regardless of the initial dislocation position, the dislocations consistently align vertically before evolving through pure vertical motion, as is especially evident in the bottom-row plots. 
This suggests that the vertical direction, parallel to the Burgers vector, is a preferential migration path. Such behavior is reminiscent of dislocation glide along a stable slip plane in conventional crystals.

\begin{figure}
    \centering
    \includegraphics[width=\textwidth]{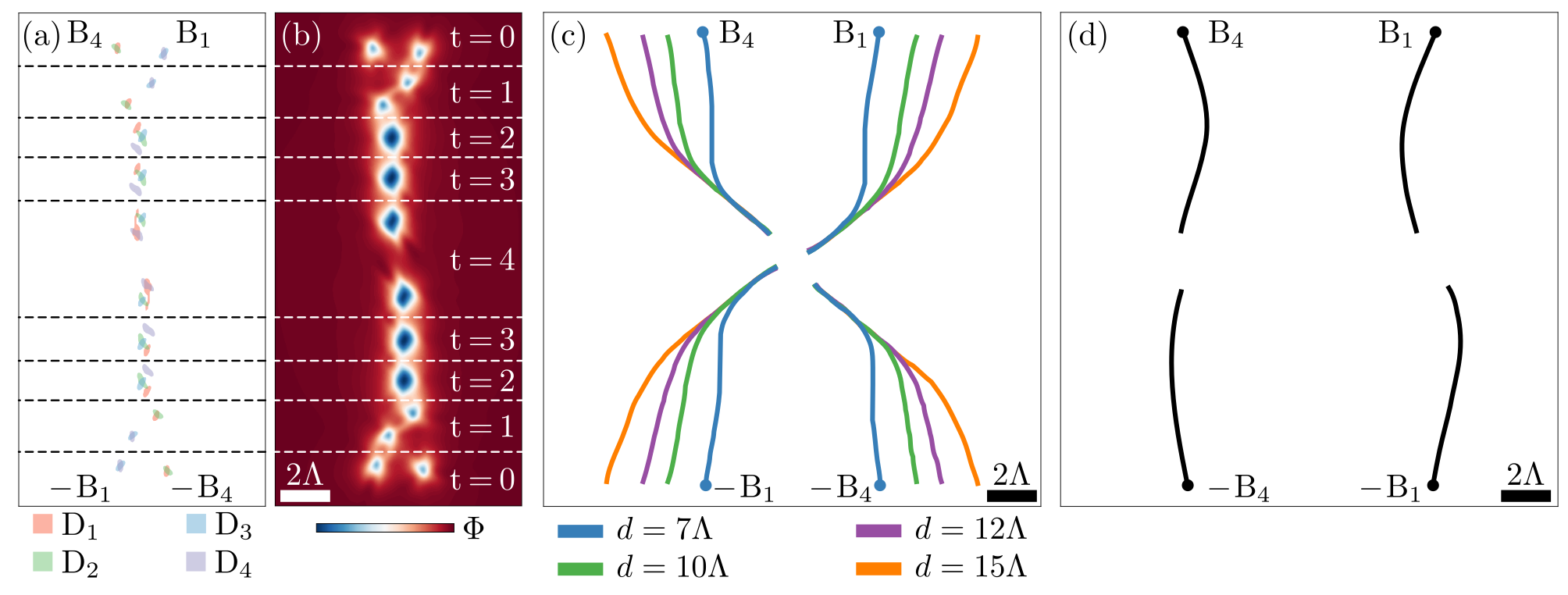}
    \caption{
    \textit{Dislocation quadrupoles.}
    (a,b) Plots of the D field and the order parameter $\Phi$ for the time evolution of two dislocation dipoles with Burgers vectors $\pm \mathbf{B}_1$ and $\pm \mathbf{B}_4$. 
    The vertical distance between the dislocations is $18\Lambda$, and the horizontal distance is $2\Lambda$. 
    Snapshots corresponding to different timesteps are labeled in units of $10^4$.  
    (c) Plot of the dislocation trajectories for larger, variable horizontal distance $d$. 
    The arrangement of the dislocations and their vertical distance are the same as panel (a).
    (d) Plot of dislocation trajectories for a configuration where the bottom dislocations are exchanged w.r.t the previous cases. The vertical distance between the dislocations is $18\Lambda$, and the horizontal distance is $10\Lambda$.  
    Simulation parameters are the same as in Fig.~\ref{fig:model}.
    }
    \label{fig:2dipoles}
\end{figure}

In Fig.~\ref{fig:panino}, we showed that the dislocation with Burgers vector $\mathbf{B}_n+\mathbf{B}_{n+2}$ has a significantly higher energy than those with Burgers vector $\mathbf{B}_n$ and $\mathbf{B}_n+\mathbf{B}_{n+1}$. 
The results from Fig.~\ref{fig:dipole_type2}(c) suggest that such a dislocation is stable, meaning that it does not split into dislocations of the type $\mathbf{B}_n$. 
To further showcase the stability of this composite dislocation, we consider two dipoles of low-energy dislocations, with Burgers vectors $\pm\mathbf{B}_1$ and $\pm\mathbf{B}_4$, positioned at $\pm(\Lambda,9\Lambda)$ and $\pm(-\Lambda,9\Lambda)$ respectively.
Figure~\ref{fig:2dipoles}(a,b) illustrates their dynamics with snapshots equally spaced in time, separated by dashed lines. The system evolves by first merging the simple dislocations into composite dislocations and only afterwards annihilating the composite dislocations. 
We remark that this behavior is not due to the initial dislocations already forming an extended dislocation: it is visible from the distribution of the topological charges in Figure~\ref{fig:2dipoles}(a) that the low-energy dislocations are well separated for over $10^4$ timesteps, and evolve similarly to the dislocations in Fig.~\ref{fig:dipole_type2}(b) before eventually merging. 
Therefore, we have shown that one high-energy dislocation $\mathbf{B}_n+\mathbf{B}_{n+2}$ can form by the reaction of two low-energy dislocations ($\mathbf{B}_n$) when the interaction of the latter is strong such to overcome the energy difference $\Delta F(\mathbf{B}_n+\mathbf{B}_{n+2})-2\Delta F(\mathbf{B}_n)$.

For the chosen distance along the $y$-axis of $18\Lambda$ in Fig.~\ref{fig:2dipoles}(a,b), we found that the merging of the initial dislocations occurs up to a distance along the $x$-axis of $6\Lambda$. 
For larger horizontal distances, the dislocations no longer merge but instead annihilate simultaneously in the center of the system. This is in agreement with the energy values shown in Fig.~\ref{fig:panino}(b), for which $2\Delta F(\mathbf{B}_n)\lesssim \Delta F(\mathbf{B}_n+\mathbf{B}_{n+2})$, i.e. $\mathbf{B}_n$ dislocations are expected to be favored over  $\mathbf{B}_n+\mathbf{B}_{n+2}$ dislocations for large distances. 
In Fig.~\ref{fig:2dipoles}(c), we show the dislocation trajectory for selected horizontal distances $d$. 
We remark that two preferential directions of movement appear, similar to what we observed in Fig.~\ref{fig:dipole_type2}(c). However, due to the presence of multiple dislocations, the preferential direction does not align with any Burgers vector, as it is the result of the collective motion of four dislocations.
Finally, in Fig.~\ref{fig:2dipoles}(d), we change the initial condition by exchanging the position of the two bottom dislocations so that dislocations with opposite Burgers vectors are aligned vertically. 
The interaction between the two dipoles is minor, even at a relatively small initial horizontal distance of $10\Lambda$. Each dislocation annihilates with its opposite, following trajectories that closely resemble those of isolated dislocation dipoles in Fig.~\ref{fig:dipole_type2}(b).

\section{Conclusions}
\label{sec:conclusions}

We have demonstrated how a mesoscale approach based on amplitude equations provides a powerful framework for analyzing, modeling, and simulating dislocations in QCs. 
By capturing the topological features of dislocations as well as their induced deformations, this method enables the identification of dislocation structures and their Burgers vectors.
Importantly, it allows for inspecting dislocations in quasicrystalline symmetry emerging from a prescribed initial condition and/or constraints.

Our analysis of rotated inclusions, strained systems, and dislocation dipoles reveals that different dislocations are expected, even when considering the relaxation of interfaces purely driven by energy minimization. 
We characterized the energetics of these dislocations and their stability in QCs featuring ten-fold symmetry. Furthermore, the investigation of dislocation dipoles shows two scenarios for stable dislocations, which evolve with a dislocation core that is either localized or delocalized. 

Beyond the specific cases considered here, the amplitude equation approach presents a promising avenue for further exploration of dislocations in QCs. 
Future work could extend these studies to other quasicrystalline systems with different symmetries and in three spatial dimensions, where dislocation structures might exhibit additional complexity. 
Moreover, incorporating dynamic aspects, such as dislocation motion under external stresses, thermal activation, and an explicit elastic relaxation timescale \cite{Heinonen2016} would provide an even more comprehensive understanding of dislocation dynamics in QCs. 
These extensions would further establish the amplitude equation framework as a versatile tool for studying mesoscale dislocation phenomena in aperiodic materials. 

We finally remark that such a description builds on the very definition of QC based on discrete sets of incommensurate reciprocal lattice vectors, which is the main input of the model in Sect.~\ref{sec:model}. Moreover, amplitudes of density waves are well-known, convenient descriptors of QCs, even proposed in seminal works, although for bulk systems only \cite{Mermin1985,Lifshitz1997}.
It is, however, in the formulation proposed in \cite{DeDonno2024} that amplitude equations have been derived to describe both bulk phases and deformations following the definition of a suitable free energy functional. Its application to dislocations, as demonstrated in this work, thus represents a significantly novel and timely methodological advance. We anticipate its broad applicability in analyzing and tailoring dislocation configurations in various systems exhibiting quasicrystalline order.

\section*{CRediT authorship contribution statement}

\textbf{Marcello De Donno}: Conceptualization, Methodology, Software, Validation, Formal analysis, Investigation, Writing - Original Draft, Writing - Review $\&$ Editing.
\textbf{Luiza Angheluta}: Conceptualization, Methodology, Resources, Writing - Review $\&$ Editing, Supervision. 
\textbf{Marco Salvalaglio}: Conceptualization, Methodology, Resources, Writing - Original Draft, Writing - Review $\&$ Editing, Supervision, Funding Acquisition. 

\section*{Declaration of competing interest}
The authors declare that they have no known competing financial interests or personal relationships that could have appeared to influence the work reported in this paper.

\section*{Acknowledgments}
We acknowledge fruitful discussion with Ken Elder. M. D. D. and M. S. acknowledge funding by the Deutsche Forschungsgemeinschaft (DFG, German Research Foundation) Project No.~447241406, and the computing time made available to them on the high-performance computer at the NHR Center of TU Dresden.

\section*{Data availability}
Data will be made available on request.

\bibliographystyle{elsarticle-num}

\end{document}